# Large-scale neuromorphic optoelectronic computing with a reconfigurable diffractive processing unit

Tiankuang Zhou[1,2,5], Xing Lin[1,2,4,#], Jiamin Wu[1,2], Yitong Chen[1,2], Hao Xie[1,2], Yipeng Li[1,2], Jintao Fan[1,2], Huaqiang Wu[3,4,7], Lu Fang[2,3,6,#] and Qionghai Dai[1,2,3,#]

[1]Department of Automation, Tsinghua University, Beijing 100084, China
[2]Institute for Brain and Cognitive Science, Tsinghua University, Beijing 100084, China
[3]Beijing National Research Center for Information Science and Technology, Tsinghua University, Beijing 100084, China
[4]Beijing Innovation Center for Future Chips, Tsinghua University, Beijing 100084, China
[5]Graduate School at Shenzhen, Tsinghua University, Shenzhen 518055, China
[6]Department of Electronic Engineering, Tsinghua University, Beijing 100084, China
[7]Institute of Microelectronics, Tsinghua University, Beijing 100084, China

# Correspondence: lin-x@tsinghua.edu.cn, fanglu@sz.tsinghua.edu.cn, qhdai@tsinghua.edu.cn

**Application-specific optical processors have been considered disruptive technologies for modern computing that can fundamentally accelerate the development of artificial intelligence (AI) by offering substantially improved computing performance. Recent advancements in optical neural network architectures for neural information processing have been applied to perform various machine learning tasks. However, the existing architectures have limited complexity and performance; and each of them requires its own dedicated design that cannot be reconfigured to switch between different neural network models for different applications after deployment. Here, we propose an optoelectronic reconfigurable computing paradigm by constructing a diffractive processing unit (DPU) that can efficiently support different neural networks and achieve a high model complexity with millions of neurons. It allocates almost all of its computational operations optically and achieves extremely high speed of data modulation and large-scale network parameter updating by dynamically programming optical modulators and photodetectors. We demonstrated the reconfiguration of the DPU to implement various diffractive feedforward and recurrent neural networks and developed a novel adaptive training approach to circumvent the system imperfections. We applied the trained networks for high-speed classifying of handwritten digit images and human action videos over benchmark datasets, and the experimental results revealed a comparable classification accuracy to the electronic computing approaches. Furthermore, our prototype system built with off-the-shelf optoelectronic components surpasses the performance of state-of-the-art graphics processing units (GPUs) by several times on computing speed and more than an order of magnitude on system energy efficiency. We believe the proposed reconfigurable DPU is a remarkable step towards high-performance neuromorphic optoelectronic computing processors that can achieve real-time dynamic architecture configurations according to software and will facilitate a broad range of AI applications, e.g., autonomous driving, robotics, and edge computing.**

Computing processors driven by electronics have evolved dramatically over the past decade, from general-purpose central processing units (CPUs)[1] to custom computing platforms, e.g., GPUs[2], FPGAs[3], and ASICs[4,5], to meet the ubiquitously increasing demand of computing resources. The progress of these silicon computing hardware platforms has greatly contributed to the resurgence of artificial intelligence (AI) by allowing the training of larger-scale and more complicated models[6,7]. We have witnessed the extensive applications of various neural computing architectures, e.g., convolutional neural networks (CNNs)[2,7], recurrent neural networks (RNNs)[8], spiking neural networks (SNNs)[9], and reservoir computing (RC)[10], in a broad range of fields. However, electronic hardware implementations have reached unsustainable performance growth as the exponential scaling of electronic transistors embodied by Moore's law approaches its physical limit, where the speed and energy are fundamentally limited by parasitic capacitance, the tunnelling effect, and crosstalk[11,12]. Therefore, despite the current dominance of electronic processors, the development of the next-generation computing modality is particularly imminent.

Optical computing uses photons instead of electrons for computation, and this process can overcome the inherent limitations of electronics and improve the energy efficiency, processing speed, and computational throughput by orders of magnitude[13,14]. Such extraordinary properties have been exploited to build application-specific optical processors for solving fundamental mathematical[15-21] and signal processing[22-25] problems with performances far beyond that of existing electronic processors. In particular, the artificial neural networks (ANNs) are one of the most promising optical computing paradigms, where the neuron functionality and its dense interconnectivity can be effectively implemented with optoelectronic devices and the nature of light propagation[26-29]. Substantial recent progress has been made in optically accelerated neural information processing to accomplish some advanced AI tasks[30-43]. Nevertheless, existing optical AI accelerators can only support a single functionality customized for the specific neural network architecture or task and cannot be adapted to diverse AI algorithms for different tasks. In addition, the model complexity of current optical neural networks (ONNs) is much lower, resulting in a large gap in model accuracy (e.g., classification accuracy) compared with that of state-of-the-art electronic ANNs. The reasons are mainly due to (1) the limited flexibility of the network design space in optics, e.g., the difficulty of integrating ideal nonlinear operations and flexibly controlling the complex data flow, and (2) the imperfections of optical systems, which cause the deviation of a model and the accumulation of computing errors in practice.

Here, we propose a reconfigurable diffractive processing unit (DPU) for large-scale neuromorphic optoelectronic computing that can be programmed to change its functionality and adapt to different types of neural network architectures. The proposed optoelectronic processor combines the advantages of optics and electronics with the availability of extremely high-bandwidth optical modulators and photodetectors; almost all of its computational operations (OPs) are effectively implemented with optical diffraction, and its programmability is controlled electronically. We demonstrate the flexible design of complex diffractive neural networks with a DPU that can support millions of neurons. The superior inference capabilities of these networks were validated by applying them for video-rate image recognition tasks with a computing speed of 270.5 TOPs/s and a system energy efficiency of 5.9 TOPs/J, which are over 9 times and an order of magnitude improvement over those of state-of-the-art GPUs[44], respectively. To effectively design and train the networks for the tasks, an in silico training solution[34,35] is first adopted for learning the neural network parameters and evaluating the model accuracy. Our diffractive feedforward neural networks surpass the classification accuracy of LeNet-4 (an error rate of 1.1%) on the MNIST handwritten digit database[45], and the constructed diffractive RNNs reach or even outperform the classification accuracy of state-of-the-art electronic computing approaches[38] on the Weizmann[46] and KTH[47] human action datasets. During the practical implementation of in silico-trained models with optoelectronic devices, we develop an adaptive training method to address the deviations of models caused by different error sources, such as alignment errors and non-ideal device characteristics. By adaptively fine-tuning the network parameters layer by layer to compensate for the system imperfections, the model deviations were successfully circumvented, and our experimental results achieved highly accurate inference performance, i.e., 96.8%, 100.0%, and 96.3% blind testing accuracies for the entire MNIST, Weizmann and KTH test datasets, respectively (see Extended Data Table 1).

**Optoelectronic implementation of the reconfigurable diffractive processing unit**

The principle of the proposed DPU is illustrated in Fig. 1a. The DPU is an optoelectronic neuromorphic processor comprising large-scale diffractive perceptron neurons and weighted optical interconnections. It represents a fundamental building block that can be programmed for establishing various types of diffractive neural networks with high model complexity and accuracy. To achieve optical information processing with diffractive neurons, the unit input data are quantized and electro-optically converted to a complex-valued optical field with an information coding (IC) module. Different input nodes are physically connected to individual neurons with the light diffractive connections (DCs), where the synaptic weights that control the strength of connections are determined by the diffractive modulation (DM) of the wavefront. Each diffractive optoelectronic neuron performs the optical field summation (OS) of its weighted inputs and generates the unit output by applying the complex activation (CA) function on the calculated optical field that naturally occurs during the photoelectric conversion.

We adopt programmable optoelectronic devices with an extremely high data throughput, i.e., on the order of gigabytes per second (GB/s), to implement the DPU that allows for high-speed neural network configurations and achieves video-rate inference capabilities (Fig. 1b). Our system in this work is designed to process large-scale visual signals that feed in images and videos; thus, a digital micromirror device (DMD) and spatial light modulator (SLM) are selected as optical modulators, and a scientific complementary metal-oxide-semiconductor (sCMOS) sensor is used as the photodetector. The DMD provides a high optical contrast to perform IC that facilitates system calibration and optical signal processing. It encodes the binarized unit inputs into the amplitude of coherent optical fields, and their phase distribution is subsequently modulated by a phase SLM in this example for implementing DM. The weighted DCs between input nodes and artificial neurons are enabled through free-space optical diffraction, where the receptive field of each neuron is determined by the amount of diffraction from the SLM plane to the sensor plane. We take advantage of the photoelectric effect characteristic of sCMOS pixels to achieve the functionality of artificial neurons, i.e., OS and CA, and generate unit outputs with high efficiency. We exploit the photoelectric effect, which measures the intensity of incident optical fields, as the CA function avoids the preparation of nonlinear optical materials and reduces the system complexity. Our system schematic and experimental setup are shown in Extended Data Fig. 1. By controlling and buffering the massively parallel optoelectronic dataflow of the unit, the DPU is allowed to be temporally multiplexed and programmed for customizing different types of ONN architectures (Fig. 1c,d,e). Since almost all of the computational operations are accomplished optically, the proposed optoelectronic AI architecture dramatically improves the computating speed and system energy efficiency compared with existing electronic neuromorphic processors (see Supplementary Table S1).

Analysing light diffraction to construct ONNs with diverse diffractive optical elements has been proven to be an effective method for integrating more optical neurons[34-37,40-43]. The DPU benefits from this capability and enables us to develop different diffractive neural networks with the following unique features: (1) programmable neural network parameters for adapting the optical system imperfections; (2) high-throughput optoelectronic devices to support massive numbers of neurons and their interconnectivity; (3) high-speed dataflow control and buffer for flexibly reconfiguring complex networks; and (4) a vast majority percentage of optical computing to empower the processor with high energy efficiency. Specifically, we demonstrate that the sequential cascading of DPU layers successfully achieves a diffractive deep neural network (DNN), i.e., $D^2NN$[34-37], as shown in Fig. 1c, where the experimental results and numerical designs for classifying the MNIST handwritten digit database were highly matched after the adaptive training. To further improve the inference capability for higher classification accuracy, we increase the model complexity of the constructed feedforward neural network by generating multiple diffractive feature maps at each hidden layer and constructing a diffractive network in network (D-NIN) architecture (Fig. 1d). In addition to the feedforward architectures, the DPU can be recurrently connected as diffractive RNNs, i.e., D-RNNs (Fig. 1e), for feeding in the temporal sequential data, and these networks are applied for high-accuracy human action video recognition on the Weizmann and KTH databases.

**Adaptive training of optoelectronic diffractive deep neural network ($D^2NN$)**

We validate the effectiveness of the proposed adaptive training approach as well as the functionality of the DPU by configuring an optoelectronic $D^2NN$ for classifying the MNIST handwritten digits (Fig. 2). It has been demonstrated that the all-optical $D^2NN$[34-37] model can classify the MNIST database with a very promising model accuracy compared with electronic computing. However, the difficulty arises from the imperfections of the physical system, e.g., layer misalignment, optical aberration, fabrication error, etc., which inevitably deteriorate the performance of an in silico-designed network model and cause discrepancies between numerical simulations and practical experiments. We show that such system error-induced model degeneration can be effectively compensated in the optoelectronic $D^2NN$ by applying the measured intermediate optical fields of DPU outputs for adaptively adjusting the network parameters (see Extended Data Fig. 2). In contrast to in situ training solutions[32,33] that seek to update the gradient directly in the system, our adaptive training approach sequentially corrects the in silico-trained model layer by layer for higher training robustness and efficiency. To implement, we first adopt an in silico electronic pre-training process[34] to simulate the network model and learn its parameters. In silico training has lower training complexity than in situ training by taking advantage of existing high-performance physical models and system parameters. The pre-trained model is then transferred to the optoelectronic D$^2$NN system by deploying the network structure and programming the SLM for individual DPU layers to write the network parameters. As the model transfer errors distort the wavefront connections of neurons in each layer, we derive the fundamental principle from adaptive optics[48,49] to sequentially compensate for the wavefront distortion in each layer and alleviate the error accumulation. For the correction of the current layer, we experimentally record its DPU outputs in situ by using the samples from the training set and adopt them as the inputs of the next layer for in silico re-training of the following diffractive layers. The error of the last layer is corrected by simply multiplying the energy distribution of output categories by the decision coefficients, e.g., 10 coefficients for 10 categories, and the optimization uses the same training schematic. The adaptive training process fine-tunes the parameters of each diffractive layer to accommodate the model transfer errors and effectively recovers the high inference accuracy.

The MNIST classification was performed by using a three-layer optoelectronic D$^2$NN that experimentally operated at an inference speed of 56 fps (see Methods). The model was in silico trained with an MNIST training set containing 55,000 images and achieved a blind testing accuracy of 97.6% on 10,000 digit images in a test set (see Methods and Extended Data Table 1). The training target was set to individually map input handwritten digits, i.e., from "0" to "9", into ten predefined regions on the output layer (i.e., layer 3), where a classification result was determined by finding the target region with the maximum optical signal (Fig. 2a). Without adaptive training, the direct transfer of the pre-trained model to our optoelectronic system dramatically decreased the recognition accuracy to 63.9% due to the accumulation of system errors layer by layer. As illustrated with the exemple digits "2" and "0" from a test set (Fig. 2a, left and middle), the layer error accumulation causes the intensity distribution of the DPU output at each layer to gradually deviate from that of the pre-trained model, and this deviation reduces the percentage of energy focusing on the target region and thus may result in an incorrect recognition category. For example, the simulation output of a digit "0" has 90.8% of its energy distribution percentage correctly focused on the zeroth region; however, this percentage decreases to 5.5% during the experiment and is lower than the value of 25.2% for the sixth output region, i.e., misclassified as a digit "6". The confusion matrix (Fig. 2c, left) summarizes the classification results of all instances in the MNIST test set, where the diagonal and non-diagonal values represent the percentages of correct and incorrect predictions for each category, respectively. The results show that transferring a direct model causes a large percentage of incorrect predictions and is especially biased for those digits such as "0", "3", "8", and "9" that achieve correct prediction percentages of less than 60%.

To circumvent the system error and improve the recognition performance, adaptive training of the constructed three-layer optoelectronic D$^2$NN was implemented with two-stage fine-tuning of the pre-trained model. Specifically, we had the option to trade-off between the experimental accuracy and training efficiency by using a full training set or a mini-training set (2% in this example), where the DPU outputs of the first and second layers were experimentally recorded during the first and second adaptive training stages, respectively. The first adaptive training stage used the experimentally measured outputs of the first layer as the inputs of the second layer and re-trained the parameters of the second and third diffractive layers in silico. Similarly, the second layer's experimental outputs were used for in silico re-training of the third diffractive layer during the second adaptive training stage. Each adaptive training stage performed in silico fine-tuning of the network parameters by initializing this process with the pre-trained model and keeping the same training settings. After each stage, the SLM phase patterns were updated accordingly with the refined parameters for adapting the system imperfections and alleviating the accumulation of system errors. With adaptive training, the intensity distributions of DPU outputs between simulations and experiments are more matched, especially at the last layer, and both of the exemple testing digits "2" and "0" are correctly categorized during the experiments (Fig. 2a, right). More exemple of correct $D^2NN$ testing results can be founded in Supplementary Video 1. The convergence plot in Fig. 2b shows that by refining the parameters of the pre-trained model with experimentally measured layer outputs, the first and second stages of adaptive training improve the system error-induced testing accuracy decrement from 82.1% to 97.4% and 84.9% to 96.3%, respectively, with a full training set (orange plots) and from 82.1% to 95.9% and 75.4% to 93.6%, respectively, with a mini-training set (yellow plots). Despite the lower accuracy, the mini-training set is more efficient, i.e., ~3 min for 20 epochs with the mini-training set compared with ~3.8 h of pre-training (blue plot) for 15 epochs with the full training set. After the adaptive training with multiplying of decision coefficients, the large model discrepancies between the simulations and experiments are overcome, and the experimental testing accuracy improves from 63.9% (green star) to 96.0% (orange star) with the full training set and to 93.9% (yellow star) with the mini-training set. As shown by the confusion matrix in Fig. 2c, right, the correct prediction rates of the categories are improved and are all larger than 93%. A histogram of the phase differences (Fig. 2d, bottom left) between the diffractive layers before (Fig. 2d, top) and after (Fig. 2d, bottom right) adaptive training reflects the fine-tuning process of the second and third diffractive layers for adapting the system error, where a large percentage of diffractive neurons have a small change in the phase modulation value.

**High-accuracy object classification with a diffractive network in networks (D-NIN-1)**

In a CNN architecture (e.g., LeNet[45]), segmenting the hidden layer into a set of feature maps with weight sharing is the critical mechanism that leads to high model performance. We demonstrate that the inference capability and model robustness of the optoelectronic $D^2NN$ can be further enhanced by designing a multi-channel diffractive hidden layer as well as its external and internal interconnectivity structure (Figs. 1d and 3a). Since the phase modulation of unit inputs in the DPU is inherently equivalent to the complex convolution operation in the frequency domain, we implement multiple diffractive feature maps at each hidden layer by stacking multiple DPU layers. Each feature map is generated with high-speed buffering and the weighted summation of DPU outputs and is set to share the same DPU layer for reducing the network parameters and achieving high-efficiency training. We term the constructed architecture a diffractive network in network, i.e., D-NIN-1, as each of the diffractive feature maps is externally weighted and connected to all feature maps of the previous layer through the shared internal connectivities of the DPU layer. In contrast to the deep electronic network in network (NIN) architecture[50] that builds micro neural networks within the receptive field of successive layers, the internal neural networks in D-NIN-1 are implemented with DPU layers that fully connect the input and output feature maps through optical diffractions. To fuse the multi-channel diffractive feature maps, a schematic of the external connections of feature maps is adopted to perform the weighted summation of DPU outputs for the input feature maps of previous layers. Given the complex internal and external neuron connectivity structure used to compute more abstract features of each hidden layer, the network outputs of D-NIN-1 are obtained after a DPU read-out layer for the final decision making.

We evaluate the performance of D-NIN-1 by constructing a three-layer architecture, as shown in Fig. 3a, and demonstrate its superior model accuracy and robustness over the three-layer $D^2NN$ (Fig. 2) on the task of MNIST classification. Instead of having a single diffractive feature map at each layer, the hidden layers of D-NIN-1 are configured with three diffractive feature maps corresponding to three DPU layers in this example (see Methods). With the same DPU settings and in silico training procedures as $D^2NN$, the D-NIN-1 model improves the blind testing accuracy from 97.6% to 98.8%, surpassing the model accuracy of the electronic CNN architecture LeNet-1. A comparisons of convergence plots between D-NIN-1 and LeNets, including the LeNet-1, LeNet-4, and LeNet-5, in Fig. 3b shows that D-NIN-1 has the highest convergence speed with respect to the training epochs (see arrow). In addition, the model performance of D-NIN-1 can be further boosted as the D-NIN-1++, which achieves an accuracy of 99.0% that surpasses the LeNet-4 by integrating a low-complexity electronic fully connected layer at the end of D-NIN-1 (see Methods). Despite the integration of electronic computational operations, optical computing is still dominant, and the ratio between optical and electronic computational operations is $1.16 \times 10^6$, which guarantees high performance in terms of computing speed and energy efficiency (see Supplementary Information).

Compared with the optoelectronic $D^2NNs$, the stacking of multiple DPU layers at each hidden layer provides higher degrees-of-freedom and robustness for fine-tuning the pre-trained model of D-NIN-1 to accommodate the system imperfections. With the programming of the optoelectronic DPU system to deploy the D-NIN-1 model, the experimental classification accuracy over the whole test dataset reaches a blind testing accuracy of 96.8% after adaptive training. The confusion matrix (Fig. 3c, left) summarizing the inference results of 10,000 testing digits shows that D-NIN-1 improves the percentage of correct predictions for all of the categories (except digits "1" and "9") compared with the $D^2NN$ and achieves values larger than 94%. We also analysed the energy distribution of ten predefined detection regions on the output layer of the inference results (Fig. 3c, right). D-NIN-1 achieves an average energy percentage of more than ~50.0% on the target region for each category, as shown in the diagonal values of the energy matrix, which demonstrates that the network successfully learns to generate the maximum energy on the correct target regions for input digits. Fig. 3d (top row) visualizes the phase modulation coefficients of each DPU after the adaptive training of the constructed three-layer D-NIN-1. The corresponding feature maps at each layer and the classification results of two exemple digits "3" and "5" are shown in the middle and bottom rows of Fig. 3d, respectively. Both of the exemple digits were misclassified in the $D^2NN$, i.e., misclassifying the digit "3" as "8" and the digit "5" as "3" due to the similarity of the digit profiles on the right and bottom parts, respectively. D-NIN-1, by contrast, can correctly categorize the two digits by gradually abstracting more feature representations of input digits for making the classification decisions. The experimental results show maximum energy distribution percentages of 32.3% and 53.2%, correctly focusing on the third and fifth target regions, respectively (Fig. 3d, last column). More example of correct D-NIN-1 testing results can be founded in the Supplementary Video 2.

**Configuring a diffractive RNNs (D-RNN) for human action recognition**

In addition to still images, the reconfigurability of the DPU allows us to construct a large-scale diffractive recurrent neural network (D-RNN) to perform high-accuracy recognition tasks for video sequences. To demonstrate its functionality, we configure a standard RNN architecture based on the recurrent connections of DPU layers and apply it for the important task of video-based human action recognition[38]. The folded and unfolded representations of the proposed D-RNN are shown in Figs. 1e and 4a, respectively, and comprise the temporal sequential connections of the input, hidden, and output layers with the shared diffractive parameters in time. The memory of input sequences is formed by generating the diffractive hidden states at different time steps. For the D-RNN hidden layer at the time step of $t$, the hidden state $h_t$ is a function of the hidden state $h_{t-1}$ at the time step of $t-1$ and the input sequence $x_t$ at the time step of $t$. We adopt an affine combination to fuse the states from these two sources, i.e., $h_t = \lambda f_1(h_{t-1}) + (1-\lambda)f_2(x_t)$, where $m_t = f_1(h_{t-1})$ denotes the memory state mapping from $h_{t-1}$; $i_t = f_2(x_t)$ denotes the input state mapping from $x_t$; and $\lambda \in (0, 1)$ is the fusing coefficient that controls the strength of the memory state with respect to the input state. The complexity of mapping functions $f_1(\cdot)$ and $f_2(\cdot)$ can be increased by using multiple DPU layers, forming a spatial deep hidden layer structure in addition to the temporal deep architecture of the D-RNN, e.g., shown in Fig. 1e. Considering the recognition speed of our system and the complexity of our tasks, both of the functions are implemented with a single DPU layer (Fig. 4a) that operates the system at a read-in speed of ~70 fps (see Methods). The last hidden state of the D-RNN, summarizing the memory of input sequences, is extracted and read out with the DPU or electronic output layer to generate the categorical output distribution for determining the action categories.

The constructed D-RNN for the task of human action recognition was evaluated on two standard benchmark databases, i.e., the Weizmann[46] and KTH[47] databases, with preprocessing (see Methods) to adapt to the network input. The Weizmann dataset, including ten categories of natural human actions (bend, jack, jump, pjump, run, side, skip, walk, wave1, and wave2), was split into 60 and 30 videos (actions) as the training and test sets, respectively, with 30~100 frames for each video sequence. For the KTH dataset, we used the first scene (150 videos) with a 16:9 database split for training and testing the system[38], where each of the video sequences had 350~600 frames belonging to one of six human action categories: boxing, handclapping, handwaving, jogging, running, and walking. The recurrent connection of the hidden layer at different time steps allows the D-RNN to process a variable sequence length of inputs. Although a longer network sequence length (larger $N$) can incorporate more frames for the recognition decision, this causes difficulties for the network in training as well as forgetting of long-term memory, i.e., the vanishing of frame information at a time step that is far from the current time step. Therefore, for each of the video sequences in the database with a length of $M$, we set $N \ll M$ and divided the sequence into numbers of sub-sequences with the same length as $N$, with which the D-RNN was trained and tested. We quantitatively evaluated the model accuracy with two metrics, i.e., the frame accuracy and video accuracy. The frame accuracy was obtained by statistically summarizing the inference results of all sub-sequences in the test set. The video accuracy was calculated based on the predicted category of each video sequence in the test set and was derived by applying the winner-takes-all strategy[38] (action category with the most votes) on the testing results of all sub-sequences in the video sequence.

After the ablation and performance analyses, the network sequence lengths were set to 3 and 5 for the Weizmann and KTH databases, respectively (see Methods). The D-RNN architecture was first evaluated by configuring the DPU read-out layer and was then in silico pre-trained with the optimal fusing coefficient of 0.2 for both the Weizmann and KTH databases; it achieved a blind testing frame accuracy of 88.9% for both databases, corresponding to a video accuracy of 100% and 94.4% for the two models, respectively. To implement the model experimentally, we performed adaptive training by fine-tuning the modulation coefficients of only the read-out layer due to the recurrent connection inherence of the D-RNN. The designed modulation coefficients of the memory, read-in, and read-out DPU layers after adaptive training are shown in Fig. 4b, where the top and bottom rows correspond to the models of the Weizmann and KTH databases, respectively. Compared with the experimental results without adaptive training, the adaptive training improved the experimental frame accuracy from 51.0% to 82.9% and the experimental video accuracy from 56.7% to 96.7% for the Weizmann database. Similarly, after the adaptive training, the experimental frame and video accuracies improved from 53.8% to 85.1% and from 55.6% to 94.4% for the KTH database, respectively. Without adaptive training, the hidden layer of the D-RNN accumulates system imperfections at different time instances, resulting in a dramatic reduction in the experimental recognition accuracy. However, the adaptive-trained DPU read-out layer was able to extract the preserved temporal information of the sub-sequences and recover the model accuracy.

We visualize the experimental testing results of all sub-sequences with the categorical voting matrix in Fig. 4c by calculating the percentage of votes for all categories in each testing video sequence, where the category with the maximum percentage of the vote represents the predicted category of a video sequence. The target testing video sequences were ranked in order with respect to the video categories so that the diagonal positions of two categorical voting matrices represent the correct predictions. The experimental results show the miscategorization of one video sequence (i.e., 25th) and three video sequences (i.e., 6th, 11th, and 15th), marked with white arrows, for the Weizmann (Fig. 4c, left) and KTH (Fig. 4c, right) databases, respectively. In addition, there are more incorrect predictions of sub-sequences among the actions with higher similarities, such as between the actions of handclapping and handwaving (category labels of 1 and 2, respectively) for the KTH database (Fig. 4c, right). Four exemple testing results are shown in Fig. 4e, including the action categories of side and jack from the Weizmann database (Fig. 4e, left) as well as handwaving and walking from the KTH database (Fig. 4e, right). At the different time steps of $t$, the D-RNN hidden layer gradually updates its output states $h_t$ by sequentially reading in the input frames $x_t$ that generate the memory of input sub-sequences (first to second-last columns of Fig. 4e, left and right). The output regions predefined during the training, i.e., ten and six regions for the Weizmann and KTH databases, respectively, each corresponds to one category, and the DPU output layer reads out the memory of input sub-sequences and correctly categorizes the four exemple sub-sequences by producing the maximum energy on the target regions (last columns of Fig. 4e, left and right). More exemple of correct sub-sequence testing results from all actions of the Weizmann and KTH databases can be founded in Supplementary Video 3 and 4, respectively.

The recognition accuracy and robustness of the D-RNN can be further enhanced, forming the D-RNN++ architecture, by transferring the trained D-RNN hidden layer and using an electronic read-out layer to replace the DPU read-out layer (Fig. 4a, right). Inspired by the memory read-out modality of reservoir computing[10,38,51], we adopt an electronic linear fully connected layer that takes the last hidden state as the input nodes and fully-connects to the output action category nodes, where the weights are effectively learned with a fast ridge regression algorithm[38,51] (see Methods). The redundancy of memories in the hidden state enables us to reduce the complexity of the electronic read-out layer, i.e., reducing the number of read-out nodes, by average pooling the last hidden state. We evaluate the experimental frame and video accuracies of D-RNN++ with respect to the number of read-out nodes, as shown in Fig. 4d. The results demonstrate that D-RNN++ achieves an experimental video accuracy of 100% and 96.3% under the optimized number of electronic read-out nodes of 2500 for the Weizmann and KTH (first scene) databases, respectively. The low-complexity implementation of the electronic read-out layer has subtle effects on the proportion of optical computing operations for the D-RNN (see Supplementary Information) and preserves its computational efficiency. Furthermore, the experimental video accuracy of D-RNN++ for categorizing the Weizmann and KTH (first scene) databases achieves comparable performance to and even outperforms the state-of-the-art electronic computing approaches[38,52,53] that have reported accuracies of 100% and 96.0%, respectively.

**Discussion**

The DPU system in this work was configured with and 8-bit phase modulation accuracy with a 16-bit accuracy of optical field measurement. We adopt a DMD to provide the binary unit input and abstract the features for recognition tasks considering its properties of high optical contrast, high speed, and easy calibration property. A higher input bit depth can be achieved by using an amplitude-SLM with additional calibrations or high-speed changing of the binary pattern multiple times within a single cycle of the DPU working flow. The system can be made more compact by using a smaller modulation pixel pitch to reduce the propagation distance for diffractive connections. Since the adopted programmable optoelectronic devices are widely used in many areas, such as microscopic imaging[49] and free-space optical communication[54,55], the proposed architecture can easily be integrated into these applications with both real-space and Fourier-space implementations[35].

The computing speed and energy efficiency of the DPU can be further improved by using optoelectronic devices with higher data throughput and speed, such as detectors using single-photon avalanche diode (SPAD) arrays[56] and SLMs using the tuneable meatasurface[57], and designing the ASICs for dataflow control and unit programmability. In addition, spatial and spectral multiplexing techniques can be incorporated. Since phase modulation does not consume much energy compared to amplitude modulation, the system can be scaled up with multiple layers of phase modulation for more flexible control of network weights while preserving its light efficiency. The DPU in this implementation was designed under a single visible wavelength of 698 nm and can be extended to multiple wavelengths by using the wavelength division multiplexing (WDM) technique. Improved processing speed and throughput will facilitate the DPU to configure feedforward and recurrent neural networks with higher model complexity and inference capability. For example, we can design more advanced internal and external connectivities for the D-NIN by encoding multiple diffractive feature maps with multiple wavelengths; a long short-term memory (LSTM) unit[8] can be incorporated for the D-RNN to avoid the forgetting of long-term memory and read-in longer sequences.

**Conclusion**

We have experimentally demonstrated a reconfigurable optoelectronic computing processor, i.e., a DPU, that can be programmed to adapt to different types of ANN architectures for large-scale optical neural information processing. Benefitting from high-bandwidth optoelectronic devices with the proposed novel adaptive training technique, our approach addressed the long-standing issues of insufficient model complexity and experimental model deviations in building the neuromorphic photonic computing systems. The constructed optoelectronic neural networks were applied to video-rate recognition tasks for handwritten digits as well as human actions on benchmark databases, and the accuracy of these networks reached and even outperformed the model accuracy of advanced electronic computing methods. Our prototype system has achieved several times higher computing speeds and more than an order of magnitude better system energy efficiency than the Tesla V100 GPU[44], and the performance could be further enhanced by using on-chip programmable optoelectronic devices[58] for system integration. In addition, the proposed adaptive training technique can serve as a universal training approach for constructing large-scale neuromorphic photonic and electronic analog[44] computing systems with high model accuracy. By effectively designing an optoelectronic computing system to fuse the complementary advantages of optics and electronics, we anticipate that the proposed approach will enable the development of more powerful optical AI accelerators as critical support for modern computing and move towards opening a new era of artificial intelligence.

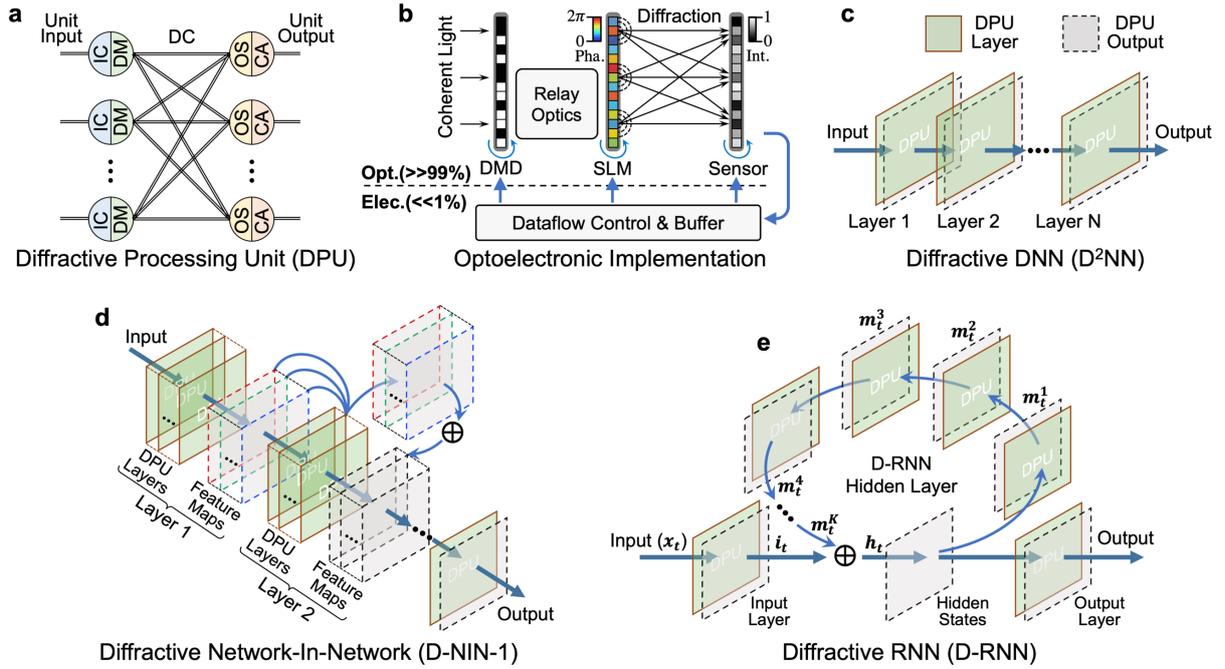

**Fig. 1. Reconfigurable diffractive optoelectronic processor. a**, A diffractive processing unit (DPU) is a large-scale perceptron-like optoelectronic computing building block that can be programmed for constructing different diffractive neural networks. IC, information coding; DM, diffractive modulation; DC, diffractive connection; OS, optical field summation; CA, complex activation. **b**, High-speed programmable optoelectronic devices were adopted to implement the DPU, i.e., a digital micromirror device (DMD) for IC of unit inputs, a phase spatial light modulator (SLM) for DM of synaptic weights, and a scientific complementary metal-oxide-semiconductor (sCMOS) sensor for OS and CA that performs the functionality of diffractive optoelectronic neurons, in this example. The artificial neurons achieve OS of weighted inputs through DCs, where the nonlinear CA function is achieved by taking advantage of the inherent photoelectric effect of sCMOS pixels that measures the intensity, i.e., square of the amplitude component, of the calculated complex optical field. Our system is able to be reconfigured to perform different machine learning tasks at video rates with extremely high energy efficiency by assigning almost all of its computations optically, while its programmability and dataflow are controlled electronically. **c,d,e**, Three different types of neural network architectures were constructed with temporal multiplexing and stacking of DPU layers for large-scale neuromorphic computing that supports millions of neurons. The high-bandwidth optical modulators (DMD and SLM) and photodetector (sCMOS sensor) empower the high data modulation speed of our system to perform video-rate recognition tasks. A diffractive deep neural network (DNN), i.e., $D^2NN^{34-37}$, is achieved with the sequential cascading of DPU layers (**a**), and its inference capability can be substantially strengthened by generating multiple diffractive feature maps on each layer and constructing a diffractive network in network (D-NIN) (**b**). The DPU layers can also be recurrently connected as a diffractive RNN (D-RNN) that can take temporal sequential data, e.g., videos, as the input and encode individual hidden states for making the final recognition decision (**c**).

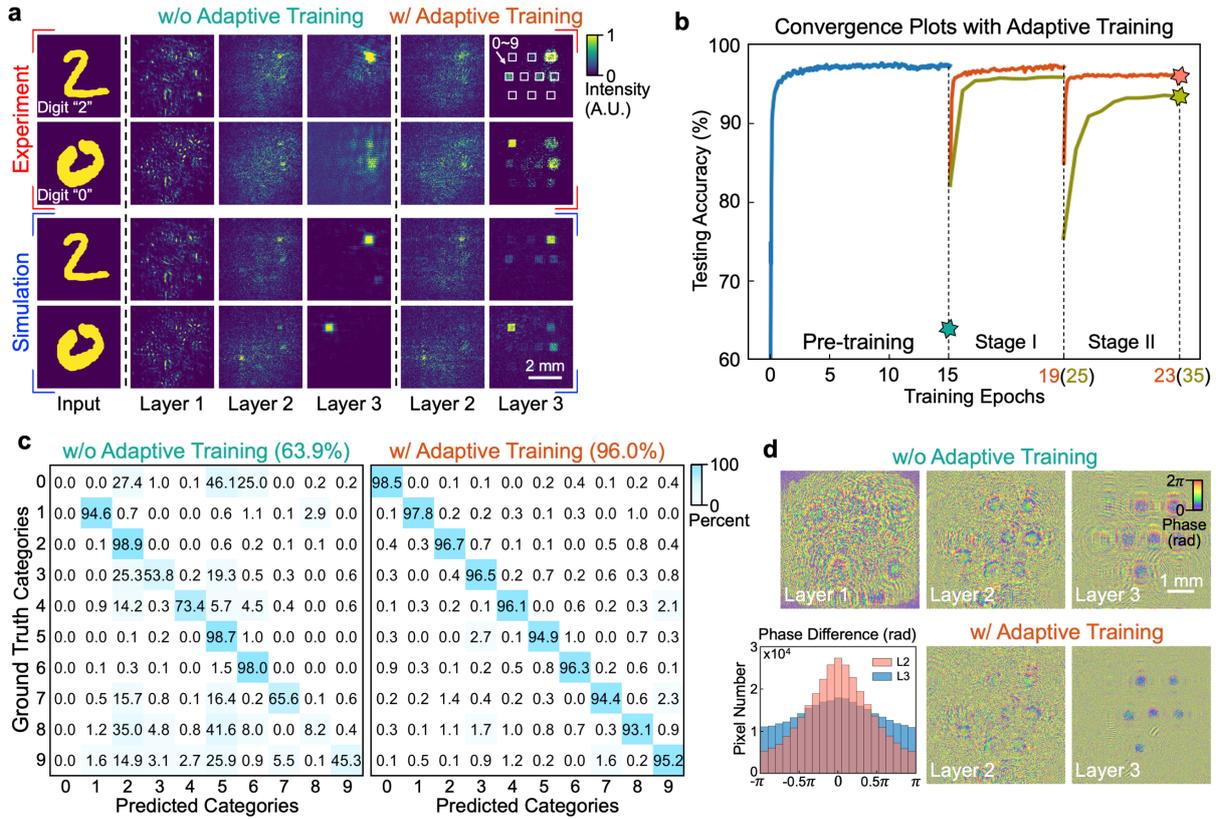

**Fig. 2. Adaptive training of the optoelectronic D²NN for handwritten digit classification.**
**a**, Experimental DPU outputs of a three-layer optoelectronic D²NN. The handwritten input digits (0~9) are classified by finding a maximum optical signal region among ten pre-defined regions at the output plane, each corresponding to one digit (from left to right, then top to bottom). With the exemple input digits "2" and "0" (left), the direct transfer of the pre-trained model to the system causes the DPU outputs of each layer to gradually deviate from those of the simulations and misclassification of the digit "0" (middle). Such an accumulation of system errors is alleviated with adaptive training, and both exemple digits "2" and "0" (right) are correctly categorized. **b**, Convergence plots (at a step size of 100 iterations) of the optoelectronic D²NN evaluated on the blind test dataset. The discrepancies between the simulations and experiments of the in silico pre-trained model dramatically decrease the blind testing accuracy to 63.9%, which is improved to 96.0% after adaptive training by using the experimentally measured DPU outputs to efficiently refine the network parameters. **c**, Confusion matrices summarizing the experimental classification results of all instances in the test set. The pre-trained model generates a high percentage of incorrect predictions for each category due to the system imperfections (left), but this limitation is effectively circumvented with adaptive training, which recovers the high prediction accuracy (right). **d**, Final designs of three diffractive layers for high-accuracy handwritten digit classification (top) and histograms of the phase differences before and after adaptive training for layer 2 and layer 3 (bottom).

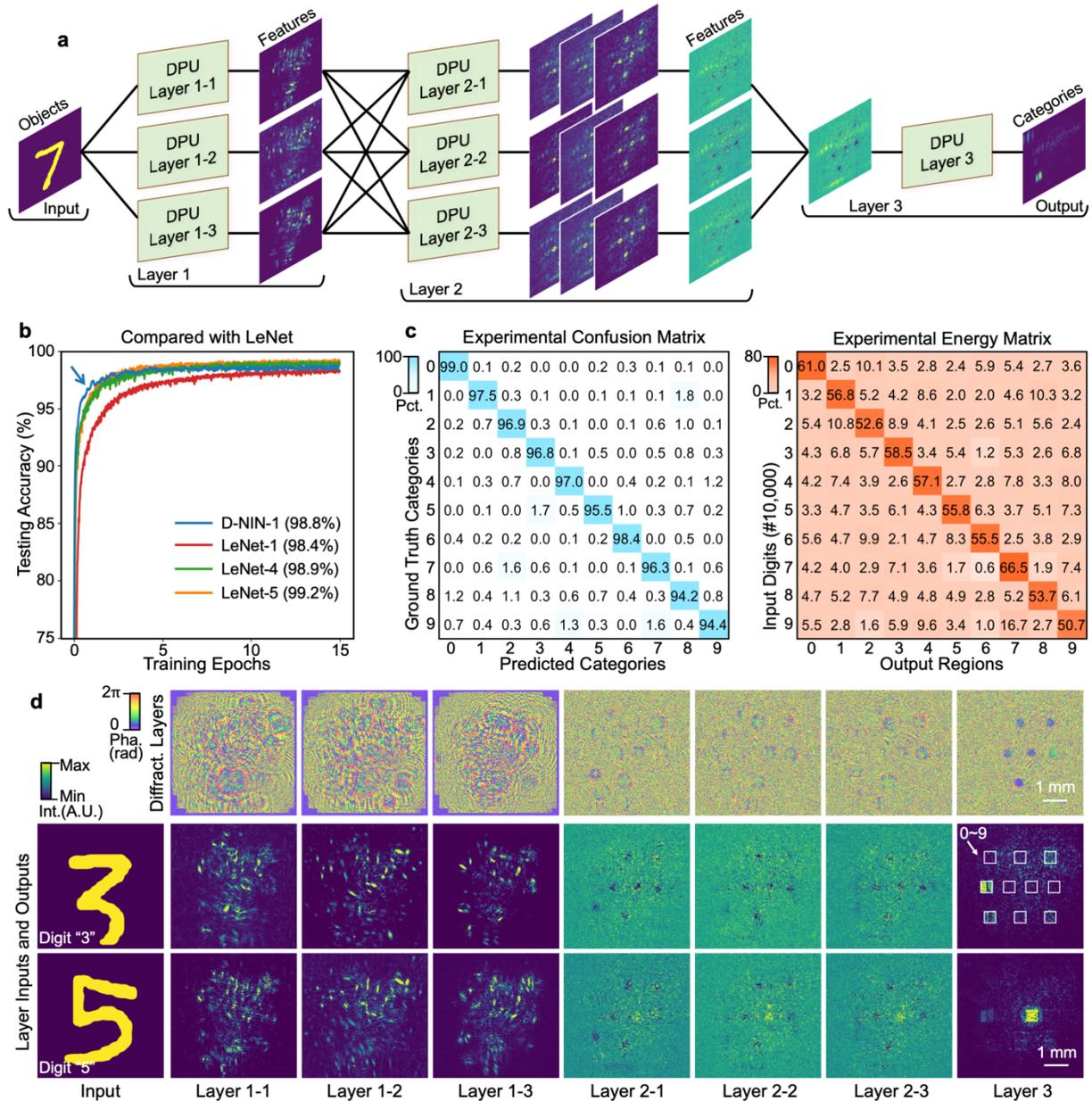

**Fig. 3. High-accuracy MNIST classification with a diffractive network in network (D-NIN).** **a**, Configuring a three-layer D-NIN-1 with DPUs. D-NIN-1 is a hierarchical interconnectivity architecture that comprises the internal and external connections of diffractive neurons. By stacking multiple DPUs at each hidden layer to generate multi-channel diffractive feature maps, the model's accuracy and the robustness of the network were substantially improved compared with those of the $D^2NN$. **b**, Convergence plots and model comparisons between D-NIN-1 and LeNet[45]. D-NIN-1 and its enhanced version (D-NIN-1++) achieve competitive classification accuracies, i.e., 98.8% and 99.0%, respectively, with respect to the electronic CNNs on the MNIST dataset. **c**, Experimental confusion and energy matrices after adaptive training. The optoelectronic system implemented with the D-NIN-1 model was blind tested with 10,000 input digits from the MNIST test set and achieved a recognition accuracy of 96.8%, and this system improved both the correct prediction rate and average energy focused on the target region. **d**, The designed diffractive layers and the layer outputs of two exemple digits "3" and "5". By learning to abstract more feature representations of input digits, D-NIN-1 correctly categorized both of the exemple digits that were misclassified by the $D^2NN$.

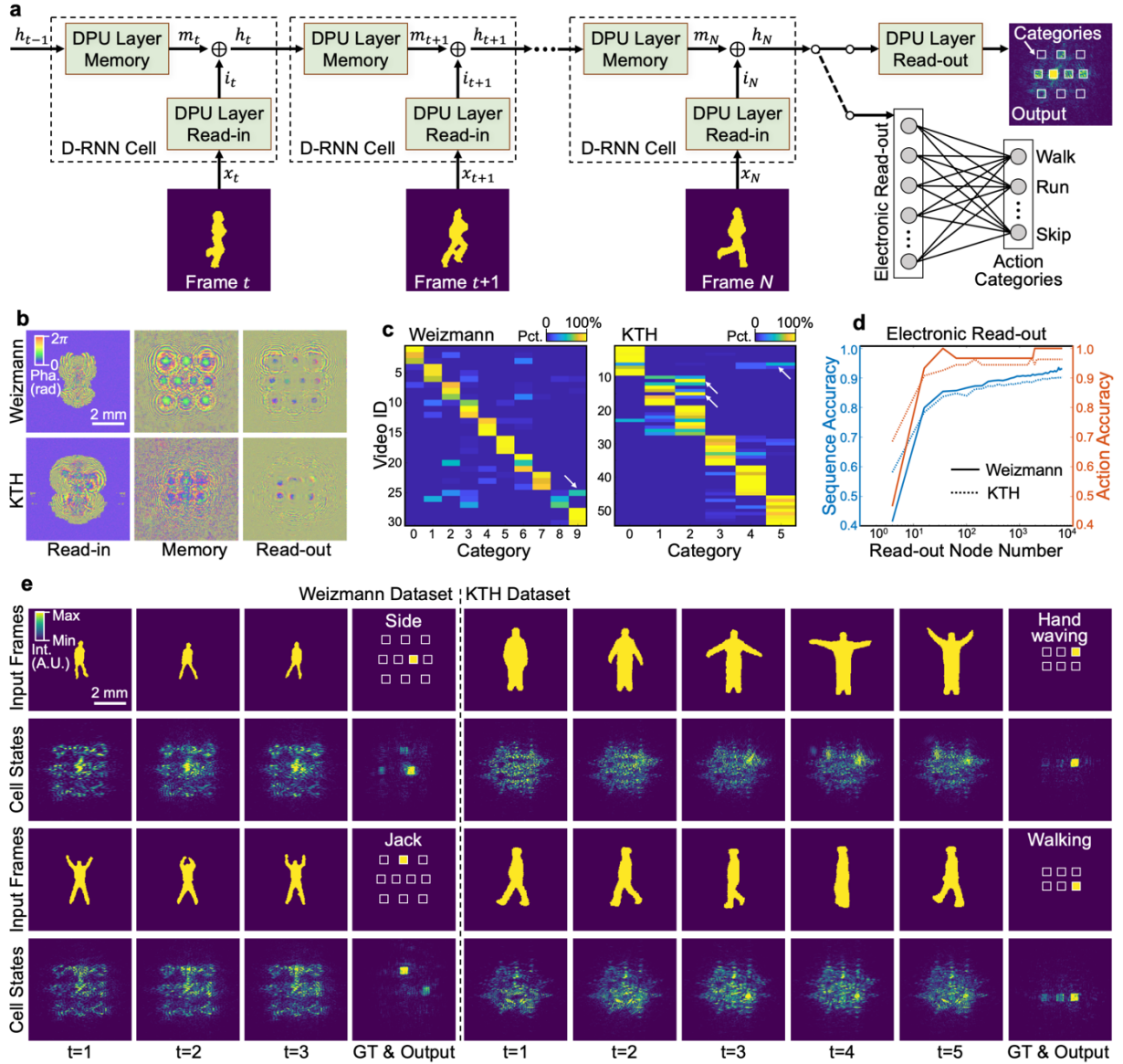

**Fig. 4. Human action recognition with a diffractive RNN (D-RNN). a**, The architecture of the unfolded D-RNN. The D-RNN uses the recurrent connection of the hidden layer for sequentially reading in the video frames at different time steps to generate the memory of inputs with the diffractive hidden states. Both the DPU and electronic output layers can be exploited to read out the memory and perform the categorization of human action videos. **b**, The designed modulation coefficients of the D-RNN after adaptive training. The performance of the D-RNN is evaluated on the Weizmann (top) and KTH (bottom) human action databases, and the network coefficients, including the memory, read-in, and read-out DPU layers, are learned for the inference tasks. **c**, The experimental inference results of all sub-sequences for the two databases. The results are visualized as a categorical voting matrix in which the category of a test video is determined by finding the maximum percentage of votes among all categories. For the Weizmann database (left), the category labels from 0 to 9 correspond to the action types of bend, jack, jump, pjump, run, side, skip, walk, wave1, and wave2, respectively. For the KTH database (right), the category labels from 0 to 5 correspond to the action types of boxing, handclapping, handwaving, jogging, running, and walking, respectively. **d**, Enhancing the performance of the D-RNN with an electronic read-out layer. The redundancy of memories

enables us to dramatically reduce the complexity of the electronic read-out layer by average pooling the last hidden state while preserving its experimental recognition accuracy. **e**, The exemple input sub-sequences, hidden states, and recognition results. Four testing sub-sequences, two from the Weizmann database (left) and two from the KTH database (right), are successfully categorized during experiments by finding the region with the maximum energy among the pre-defined regions at the output plane, with each corresponding to one category label (sequentially labelled from left to right, then top to bottom).

**Extended Data Table 1. Benchmark performance of DPU-constructed DNN architectures**

| Network Architecture Types | | Model Accuracy (%) | Experimental Accuracy (%) | Computing Speed[a] (TOPs/s) | Computing Energy Efficiency[a] (TOPs/J) | System Energy Efficiency[a] (TOPs/J) |
|---|---|---|---|---|---|---|
| $D^2NN$ | | 97.6% | 96.0% | 133.4 | 2446.1 | 2.889 |
| D-NIN-1(++) | | 99.0% | 96.8% | 133.4 | ~1514.0 | 2.887 |
| D-RNN | Weizm. | 100.0% | 96.7% | 270.5 | 2548.0 | 5.855 |
| D-RNN | KTH | 94.4% | 94.4% | 270.5 | 2316.4 | 5.854 |
| D-RNN++ | Weizm. | – | 100.0% | 270.5 | 2522.3 | 5.855 |
| D-RNN++ | KTH | – | 96.3% | 270.5 | 2281.6 | 5.853 |

([a]see Supplementary Information for the detailed calculations)